\documentclass[12pt]{article}
\pdfoutput=1
\usepackage[colorlinks=true,urlcolor=blue,citecolor=red,linkcolor=blue,bookmarks=false]{hyperref}

\usepackage{pdfpages}

\usepackage{fullpage}
\usepackage{color,graphicx}

\usepackage{amsmath,verbatim,amssymb,epsfig}
\usepackage{smartref,color}
\usepackage{longtable}
\usepackage{tabularx}
\usepackage{rotating}
\usepackage{graphicx}
\usepackage{subfig}
\usepackage{setspace}
\usepackage{booktabs}
\usepackage{amsthm}
\usepackage{datetime}

\usepackage{grffile}
\usepackage{arydshln}

\usepackage{subfig}
\usepackage{graphicx}
\def\boxit#1{\vbox{\hrule\hbox{\vrule\kern6pt\vbox{\kern6pt#1\kern6pt}\kern6pt\vrule}\hrule}}

\begin{document}
	\baselineskip=24pt
	\begin{center}
		{\Large \bf  Statistical Properties of the Keyboard Design with Extension to Drug-Combination Trials}
	\end{center}
	
	\vspace{2mm}
	\begin{center}
			Haitao Pan$^1$, Ruitao Lin$^2$ and Ying Yuan$^{2,*}$  \\

	$^1$ St. Jude Children's Research Hospital, Memphis, TN \\
	$^2$ The University of Texas MD Anderson Cancer Center, Houston, Texas \\
	$^*$ Author for Correspondence (\href{mailto:yyuan@mdanderson.org}{yyuan@mdanderson.org})
	\end{center}
	\bigskip
	
\emph{\textbf{Abstract}}: The keyboard design is a novel phase I  dose-finding method that is simple and has good operating characteristics.  This paper studies theoretical properties of the keyboard design, including the optimality of its decision rules, coherence in dose transition,  and convergence to the target dose. Establishing these theoretical properties explains the mechanism of the design and  provides assurance to practitioners regarding the behavior of the keyboard design.  We further extend the keyboard design to dual-agent dose-finding trials, which inherit the same statistical properties and simplicity as the single-agent keyboard design. Extensive simulations are conducted to evaluate the performance of the proposed keyboard drug-combination design using a novel, random two-dimensional dose--toxicity scenario generating algorithm. The simulation results confirm the desirable and competitive operating characteristics of the keyboard design as established by the theoretical study. An R Shiny application is developed to facilitate implementing the keyboard combination design in practice.

\par\vspace{3mm}

\newpage
\section{Introduction}
\label{introduction}
The objective of phase I clinical trials is to identify the maximum tolerated dose (MTD) of a new drug, which is defined as the dose with the dose-limiting toxicity (DLT) probability  closest to the target toxicity rate. Conventionally, phase I dose-finding designs are generally classified as algorithm-based and model-based \cite{iasonos2008comprehensive,le2009dose}. Algorithm-based designs implement simple, prespecified rules to guide dose escalation and de-escalation. Examples include the 3+3 design, accelerated titration design \cite{titration}, and the biased-coin design \cite{durham1997random} and its variations \cite{ivanova2003improved,stylianou2004accelerated}. The algorithm-based designs are easy to implement, but usually have unsatisfactory operating characteristics and lack theoretical support \cite{iasonos2008comprehensive,le2009dose}. Model-based designs that utilize some parametric dose--toxicity
models to make dose escalation/de-escalation decisions  have been proposed to improve upon the performance of algorithm-based designs. The most well-known example is the continual reassessment method (CRM) \cite{crm1990}, which begins with a prior guess of the dose--toxicity curve and then continuously updates the estimate of the curve based on the accruing toxicity outcomes from patients in the trial to guide the dose assignment. Various extensions of the CRM have been proposed, including dose escalation with overdose control \cite{babb1998cancer}, time-to-event CRM \cite{Cheung2000},  Bayesian model averaging CRM \cite{yin2009bayesian}, partial order CRM \cite{wages2011continual},  and bivariate CRM \cite{Braun2002bicrm}. 
For a comprehensive review of the CRM and related methods, see the book by Cheung \cite{cheung-book}. Compared to algorithm-based designs, model-based designs typically have superior operating characteristics. However, because model-based designs require repeated model fitting and estimation, they are more complicated to implement and many practitioners view the decisions of the model-based designs as coming from a ``black box".

A class of new designs, known as ``model-assisted" designs, has drawn substantial attention from practitioners because of its simplicity and desirable performance. Similar to model-based designs, these designs utilize probability models for efficient decision making, but  their dose assignment rules can be pre-tabulated before the onset of the trial in a fashion similar to the assignment rules of the algorithm-based designs. Examples of model-assisted designs include the modified toxicity probability interval (mTPI) design \cite{mTPI}, Bayesian optimal interval (BOIN) design \cite{BOIN}  and keyboard design \cite{Keyboard}.  In particular, the keyboard design is a seamless improvement of the mTPI design that addresses the latter's overdosing  issue. 
The keyboard design keeps the simplicity of the mTPI design, but has higher accuracy to identify the MTD  and substantially better overdose control \cite{Keyboard}.

The objective of this paper is to investigate the statistical properties of the keyboard design, including optimality, coherence and consistency, which are not studied in the introductory paper\cite{Keyboard}. Establishing these statistical properties is not only of theoretical interest, but also of practical importance because it assures practitioners that the design has certain guaranteed desirable behaviors.  Specifically, we show that the decisions of the  keyboard design are optimal under the ``0-1'' loss function; the design is also long-memory coherent in the sense that it never escalates (or de-escalates) the dose if the observed toxicity rate at the current dose is higher (or lower) than the target toxicity rate; last, the keyboard design is  consistent in that its dose assignment converges to the target dose under large samples. 

Armed with sound statistical properties and competitive operating characteristics, we further extend the keyboard design to handle drug-combination trials. Numerous designs have been proposed for finding the MTD for drug-combination trials: a design based on the order of the restricted inference \cite{conaway}, a six-parameter model with continuous doses \cite{thall2003dose}, a copula-type regression model \cite{yin2009copula},  latent contingency tables \cite{yin2009latent}, a Bayesian hierarchical model  \cite{braun2010hierarchical}, sequential dose-finding strategy \cite{yuan2008,zhang2016}, partial ordering CRM \cite{wages2011continual}, adaptive randomization \cite{yuan-seamless}, logistic regression model \cite{riviere2015bayesian}, change-point model with molecularly targeted agents \cite{cai}, another Bayesian optimal interval design \cite{lin2017}, and Bayesian data augmentation for late-onset toxicity \cite{liu-ning}, among others. Unlike most existing drug-combination designs, the proposed keyboard combination design enjoys the same statistical properties, while retaining the same transparency and ease of implementation as the single-agent keyboard design. A simulation study based on random and objective scenarios shows that the keyboard combination designs have very competitive performance compared to that of some existing, more complicated designs. To facilitate the use of the designs, we also develop an R Shiny application for implementing the keyboard combination designs that will be freely available at \url{http://www.trialdesign.org}.

The rest of this paper is organized as follows. In Section \ref{method}, we review the keyboard design and study its statistical properties. In Section \ref{combination}, we extend the keyboard design to two-agent combination trials. In Section \ref{simulation}, we propose a novel algorithm to generate random two-dimensional dose--toxicity scenarios, and conduct extensive numerical studies to investigate the operating characteristics of the keyboard combination design based on the randomly generated scenarios. In Section \ref{web}, we illustrate how to use the R Shiny application to design a drug combination trial. We provide some conclusions in Section \ref{conclusion}.
 
\section{The keyboard design and its statistical properties}
\label{method}
\subsection{The keyboard design}
\label{method1}

\begin{figure}
	\centering
	\includegraphics[scale=0.4]{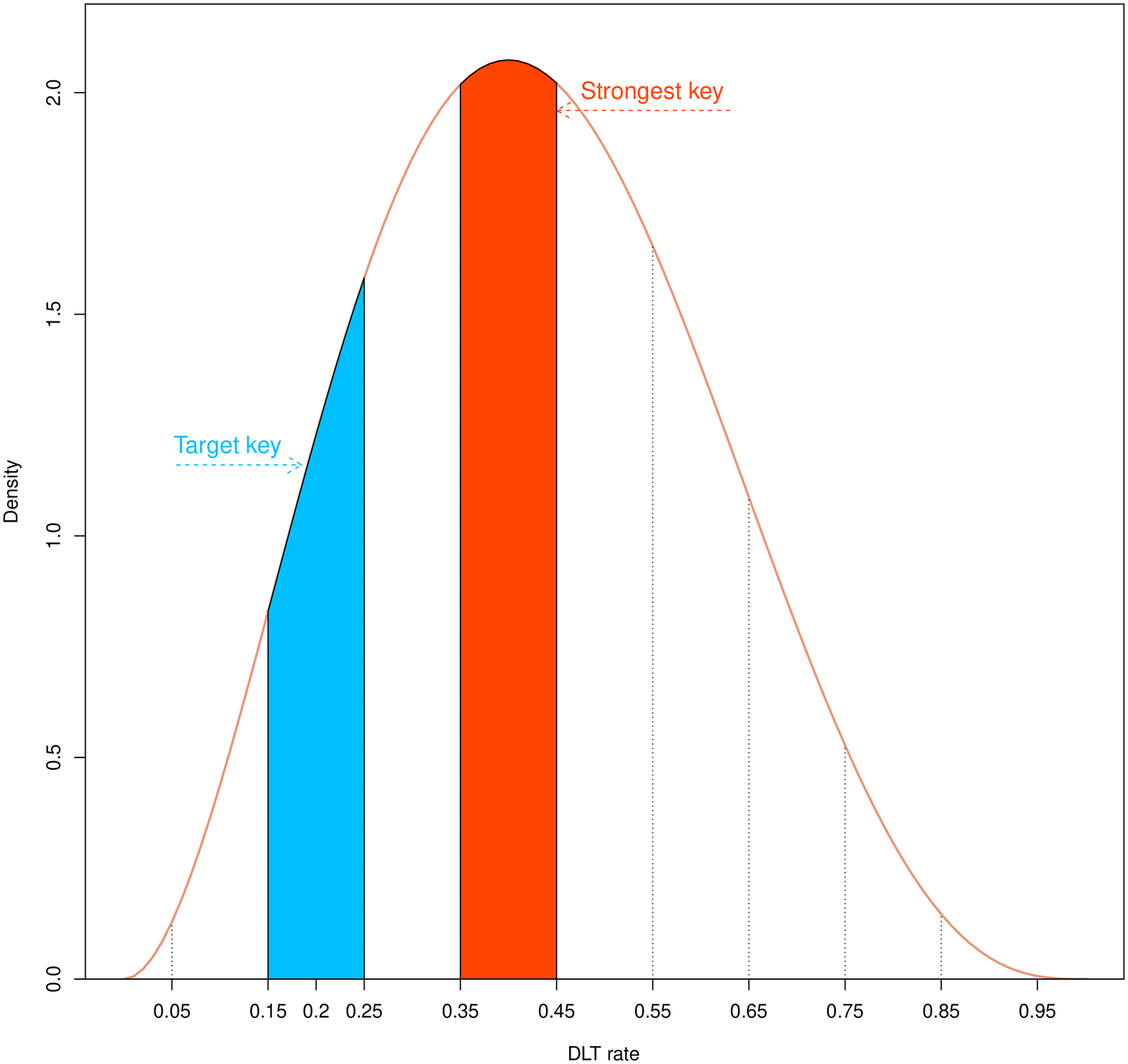}
	\caption{An example based on the observed data $(y_d,n_d)=(2,5)$ for the target toxicity rate of 0.2 to demonstrate the target key $\mathcal{I}_{\rm max}$ and the strongest keys $\mathcal{I}_{\rm target}$. In this example, the strongest key (i.e., $(0.35, 0.45)$)  lies on the right side of the target key (i.e., $(0.15,0.25)$), so dose escalation is warranted. }
	\label{fig:Keyboard_algorithm}
\end{figure}

The keyboard design is a Bayesian dose-finding design originally proposed for phase I single-agent  trials. Let $\phi$ denote the target toxicity rate specified by an investigator, and $p_d\in (0,1)$ be the toxicity probability of the dose level $d$. The keyboard design starts by specifying a target toxicity interval  ${\cal I}_{\rm target}= (\phi-\epsilon_1,\phi+\epsilon_2)$ (referred to as the target key), where $\epsilon_1$ and $\epsilon_2$ denote tolerable deviations from $\phi$ such that any dose with a toxicity probability within that interval can be practically viewed as the MTD.  Then, the keyboard design populates this interval toward both sides of the target key, forming a series of equally-wide keys  that span the range of 0 to 1 (see Figure \ref{fig:Keyboard_algorithm}).  For example, given $\phi = 0.2$, the target key may be defined as $(0.15, 0.25)$ with $\epsilon_1=\epsilon_2=0.05$. Then, one key of width 0.1 is formed on the left side of the target key, i.e., (0.05, 0.15), and seven keys of width 0.1, i.e., $(0.25, 0.35), \cdots, (0.85, 0.95)$, are formed on the right side of the target key.  We denote the resulting intervals/keys as ${\cal I}_1, \cdots, {\cal I}_K$. In some cases, the DLT probability values at the two ends (e.g., $<0.05$ or $>0.95$ in the example) may not be covered by
the keys because they are not wide enough to form a key of width 0.1. As explained in Yan et al.\cite{Keyboard}, ignoring these ``residual" DLT probabilities at the two ends does not pose any issue for decision making.

Assume that at a decision time, $n_d$ patients have been treated at the current dose  $d$, and $y_d$ of them have experienced DLT. A standard beta-binomial model is posited,
\begin{align}
	y_d\mid p_d &\sim \text{Binomial}(p_d, n_d),  \nonumber \\
	p_d &\sim \text{Unif}(0,1), \nonumber
\end{align}
where ${\rm Unif}(0,1)$ serves as a non-informative prior 
distribution for $p_d$. Given the observed data $D_d=(n_d, y_d)$, the posterior distribution of $p_d$ follows a beta distribution,
$p_d|D_d \sim \text{Beta}(y_d+1,n_d-y_d+1).$ To make the dose assignment, the keyboard design identifies the strongest  key ${\cal I}_{\rm max}$ that has the largest posterior probability, i.e., 
$${\cal I}_{\max}  = \underset{k\in\{1,\ldots,K\}}{\rm argmax} \{ {\rm Pr}( p_d \in {\cal I}_k \,|\, D_d) \}.$$
In other words, the strongest key 
${\cal I}_{\rm max}$ represents the interval where the true value of $p_d$ is most likely located. Graphically, the strongest key is the one with the largest area under the posterior distribution curve of $p_d$ (see Figure \ref{fig:Keyboard_algorithm}). 
By continuously identifying the strongest  key ${\cal I}_{\rm max}$ after each cohort, the dose-assignment rule for the next new cohort is described as follows:
\begin{itemize}
\item If the strongest key is on the left side of the target key (denoted by ${\cal I}_{\text{max}} \prec {\cal I}_{\text{\rm target}}$), 
escalate the dose to the next higher level;
\item  If the strongest key is the target key (denoted by  ${\cal I}_{\text{max}}\equiv {\cal I}_{\text{\rm target}}$), 
retain the current dose level and stay at that level;
\item If the strongest key is on the right side of the target key (denoted by  ${\cal I}_{\text{max}} \succ {\cal I}_{\text{\rm target}}$),  
de-escalate the dose to the next lower level. 
\end{itemize}
Because the decision of dose escalation and de-escalation only depends on the local data at the current dose, the decision rule of the keyboard design can be pre-tabulated, making the design transparent and easy to implement. Table \ref{escrules} shows the dose escalation and de-escalation rule for targets $\phi=0.2$ and 0.3.

\subsection{Statistical properties}
\label{property}
In this section, we study the statistical properties of the keyboard design.  Let $a\in \{\mathcal{E},\mathcal{R},\mathcal{D}\}$ denote the three possible decisions, where
$\mathcal{E},\mathcal{R}, $ and $\mathcal{D}$ denote escalating, retaining and de-escalating the dose, respectively, and let $k^*$ be the index of the key to which $p_d$ belongs. 
As the keyboard design is consisted of two sequential actions (i.e., identify the strongest key first and then make dose-assignment decisions according to the strongest key), under the Bayesian decision theoretic framework, we define a ``0-1'' loss function as follows, 
\[
\mathcal{L}(a,\hat{k}\mid\mathcal{I}_{k^{*}})=1-\ell_{1}(\hat{k}\mid\mathcal{I}_{k^{*}})\ell_{2}(a\mid\hat{k},\mathcal{I}_{k^{*}}),
\]
where $\ell_{1}(\hat{k}\mid\mathcal{I}_{k^{*}})$ is the loss function corresponding to the action of identifying the strongest key, given by
\[
\ell_{1}(\hat{k}\mid\mathcal{I}_{k^{*}})=1\{\hat{k}=k^{*}\}=\begin{cases}
1, & \hat{k}=k^{*}\\
0, & \hat{k}\not=k^{*}
\end{cases}
\]
with $1\{\cdot\}$ being the indicator function,  and $\ell_{2}(a\mid \hat{k},\mathcal{I}_{k^{*}})$ is the loss function corresponding to the action of making dose-assignment decisions according to the strongest key, given by
\begin{eqnarray*}
\ell_{2}(a\mid \hat{k},\mathcal{I}_{k^{*}}) & = & 1\{a=\mathcal{E})1\{\mathcal{I}_{\hat{k}}\prec\mathcal{I}_{{\rm target}}\}+1\{a=\mathcal{R})1\{\mathcal{I}_{\hat{k}}\equiv\mathcal{I}_{{\rm target}}\}+1\{a=\mathcal{D})1\{\mathcal{I}_{\hat{k}}\succ\mathcal{I}_{{\rm target}}\}\\
 & = & \begin{cases}
1, & a=\mathcal{E} \mbox{ and } \mathcal{I}_{\hat{k}}\prec\mathcal{I}_{{\rm target}}\\
1, & a=\mathcal{R} \mbox{ and } \mathcal{I}_{\hat{k}}\equiv\mathcal{I}_{{\rm target}}\\
1, & a=\mathcal{D} \mbox{ and } \mathcal{I}_{\hat{k}}\succ\mathcal{I}_{{\rm target}}\\
0, & \text{otherwise}
\end{cases}.
\end{eqnarray*}
Since both $\ell_{1}(\hat{k}\mid\mathcal{I}_{k^{*}})$ and $\ell_{2}(a\mid \hat{k},\mathcal{I}_{k^{*}}) $ are ``0-1'' functions, $\mathcal{L}(a,\hat{k}\mid\mathcal{I}_{k^{*}})$ is also a ``0-1'' loss function. 
Given the prior $p_d\sim{\rm Unif}(0,1)$, it is easy to obtain that  the prior model probabilities of $\mathcal{I}_k$ are equal and that
under each $\mathcal{I}_k$, $p_d$ follows  a  uniform prior distribution with the support $\mathcal{I}_k$. The expected loss function is then 
\begin{eqnarray*}
\mathcal{L}(a,\hat{k}\mid D_{d}) & = & \sum_{k=1}^{K}\mathcal{L}(a,\hat{k}\mid \mathcal{I}_{k})\Pr(p_{d}\in\mathcal{I}_{k}\mid D_{d})\\
 & = & \sum_{k=1}^{K}\{1-\ell_{1}(\hat{k}\mid\mathcal{I}_{k})\ell_{2}(a\mid\hat{k},\mathcal{I}_{k})\}\Pr(p_{d}\in\mathcal{I}_{k}\mid D_{d})\\
 & = & 1-\sum_{k=1}^{K}\ell_{1}(\hat{k}\mid\mathcal{I}_{k})\ell_{2}(a\mid\hat{k},\mathcal{I}_{k})\Pr(p_{d}\in\mathcal{I}_{k}\mid D_{d})\\
 & = & 1-\sum_{k=1}^{K}1\{\hat{k}=k\}\ell_{2}(a\mid\hat{k},\mathcal{I}_{k})\Pr(p_{d}\in\mathcal{I}_{k}\mid D_{d})
\end{eqnarray*}
Apparently, $\mathcal{L}(a,\hat{k}\mid\mathcal{D}_{d})$ is minimized if and only if $\mathcal{I}_{\hat{k}}={\cal I}_{\max}  = \underset{k\in\{1,\ldots,K\}}{\rm argmax} \{ {\rm Pr}( p_d \in {\cal I}_k \,|\, D_d) \}$ and $a$ is taken as the correct action with respect to the location of ${\cal I}_{\rm max}$. Therefore, we conclude that the  decision rules of the keyboard designs correspond to an optimal rule.

\textbf{Theorem 1 (Optimality).} The dose escalation and de-escalation rule in the keyboard designs is optimal in the sense that it minimizes the posterior expected ``0-1'' loss function $\mathcal{L}(a,\hat{k}\mid D_{d})$. 
 
Cheung \cite{cheung-book} introduced the concept of coherence  and defined it as a design property by which dose escalation (or de-escalation) is prohibited when the observed toxicity rate in the most recently treated cohort is larger (or smaller) than the target toxicity rate. Liu and Yuan \cite{BOIN} extended that concept and defined two different types of coherence: short-memory coherence and long-memory coherence. They referred to the coherence proposed by Cheung as short-memory coherence because it concerns the observations from only the most recently treated cohort, ignoring the observations from the cohorts that were previously treated. Long-memory coherence is defined as a design property by which dose escalation (or de-escalation) is prohibited when the observed toxicity rate in the accumulative cohorts at the current dose is larger (or smaller) than the target toxicity rate. From a practical viewpoint, long-memory coherence is more relevant because when clinicians determine whether a dose assignment is practically plausible, they almost always base their decision on the toxicity data that have accumulated from all patients, rather than only the most recent cohort (which rarely includes more than three patients) treated at that dose. In addition, phase I trial patients are very heterogenous and the observations from a cohort ($\le$ 3 patients) are highly variable. Thus, it can be undesirable to evaluate whether dose escalation or de-escalation is appropriate based on only a single cohort. For example, suppose the target DLT rate $\phi=0.3$ and at the current dose, one of the latest enrolled 3 patients experienced DLT but none of 6 patients previously treated at the same dose had DLT. As the overall observed DLT rate at the current dose is 1/9, escalating the dose should not be regarded as an inappropriate action, although it violates the short-memory coherence.  It can be shown that the keyboard design is long-memory coherent. 

\textbf{Theorem 2 (Coherence).} The keyboard design is long-memory coherent in the sense that the probability of escalating (or de-escalating) the dose if  the observed toxicity rate in the accumulative cohorts at the current dose is greater (or less) than the target toxicity rate, that is, $\Pr(\mathcal{E}\mid\hat{p}_{d}>\phi,D_{d})=0$, and
$\Pr(\mathcal{D}\mid\hat{p}_{d}<\phi,D_{d})=0$.

The proof  of Theorem 2 is provided in Appendix \ref{coherence}, which relies on an equivalent form of the keyboard designs (as shown in Lemma \ref{lemma}). 
However,  since the keyboard designs integrate all the accumulative data at the current dose when making decisions and does not allow dose skipping, it is not necessarily short-memory coherent. 
\smallskip

Coherence is a finite-sample property. Theorem 3 shows that the keyboard design has a desirable large-sample convergence property.  

\textbf{Theorem 3 (Convergence).} As the number of treated patients goes to infinity, the dose assignment  of the keyboard design converges almost surely to the dose level $d$ with $p_{d}\in {\cal I}_{\text{\rm target}}$. 

The proof of Theorem 3 is given in Appendix \ref{convergence}.  Although a good large-sample property does not guarantee that a design will perform well in the small sample sizes of phase I clinical trials, it provides a basic requirement for the design to perform well. If a design does not converge to the target dose under large samples, it most likely will not perform well when finding the target dose in small sample sizes. 

 \section{Drug-combination keyboard  designs}
\label{combination}
We now extend the keyboard design from single-agent trials to  combination trials with $J$ levels of drug $A$ and $K$ levels of drug $B$. Let $p_{jk}$ denote the toxicity probability of the combination of dose level $j$ of drug $A$ and dose level $k$ of drug B, denoted as $(j,k)$, $1\leq j \leq J$, $1\leq k\leq K$. Let $n_{jk}$ and $y_{jk}$ respectively denote the number of patients treated and the number of patients who have experienced toxicity at  $d=(j, k)$.
As the decision of dose escalation/de-escalation of the single-agent keyboard design is solely based on the local data at the current dose, its dose transition rule (as shown in Table \ref{escrules}) can be directly applied to drug-combination trials. A single-agent trial using the keyboard design can be summarized in the following three steps. 
\begin{itemize}
\item[Step 1.] The trial starts by treating the first cohort of patients at the lowest dose combination (1, 1). 
\item[Step 2.] At the current dose combination $(j, k)$, given the observed data $D_{jk} =(n_{jk}, y_{jk})$, we identify the strongest key ${\cal I}_{\text{max}}$
based on the posterior distribution of $p_{j, k}$:

If ${\cal I}_{\text{max}} \prec {\cal I}_{\text{\rm target}}$, we escalate the dose; 

if ${\cal I}_{\text{max}} \succ {\cal I}_{\text{\rm target}}$, we de-escalate the dose; 

otherwise, if  ${\cal I}_{\text{max}} \equiv {\cal I}_{\text{\rm target}}$, we stay at the current dose. 

\end{itemize}

\begin{itemize}
	\item[Step 3.] The process continues until the prespecified maximum sample size is achieved.
\end{itemize}

For combination trials, the difficulty is that when we decide to escalate (de-escalate) the dose, there is more than one option: we can escalate (de-escalate) either the dose level of drug $A$, the dose level of drug $B$, or both simultaneously. To address this issue, we define admissible dose escalation and de-escalation sets  $\mathcal{A}_{E_1}=\{(j+1,k),(j,k+1)\}$ and $\mathcal{A}_{D_1}=\{(j-1,k),(j,k-1)\}$, which exclude diagonal dose movements. In addition, we define admissible dose escalation and de-escalation sets that include diagonal dose movements, namely, $\mathcal{A}_{E_2}=\{(j+1,k),(j,k+1),(j+1,k+1)\}$ and $\mathcal{A}_{D_2}=\{(j-1,k),(j,k-1),(j-1,k-1)\}$.  Five dose assignment algorithms for drug-combination trials have been considered.

\noindent
\textbf{Algorithm 1}: Fixed non-diagonal escalation/de-escalation (\textbf{key1})
	\begin{itemize}
		\item Escalation: escalate  to the dose combination that belongs to $\mathcal{A}_{E_1}$ and has the highest value of $\Pr (p_{jk}\in {\cal I}_{\rm target}|D_{jk})$.
		\item De-escalation: de-escalate to the dose combination that belongs to $\mathcal{A}_{D_1}$ and has the highest value of $\Pr (p_{jk}\in {\cal I}_{\rm target}|D_{jk})$.
			\end{itemize}
\textbf{Algorithm 2}:	Fixed non-diagonal escalation and  diagonal de-escalation  (\textbf{key2})
\begin{itemize}
		\item Escalation: escalate  to the dose combination that belongs to $\mathcal{A}_{E_1}$ and has the highest value of $\Pr (p_{jk}\in {\cal I}_{\rm target}|D_{jk})$.
       \item De-escalation: de-escalate to the dose combination that belongs to $\mathcal{A}_{D_2}$ and has the highest value of $\Pr (p_{jk}\in {\cal I}_{\rm target}|D_{jk})$.
\end{itemize}	
\textbf{Algorithm 3}: Fixed diagonal escalation/de-escalation (\textbf{key3})
\begin{itemize}
	\item Escalation: escalate  to the dose combination that belongs to $\mathcal{A}_{E_2}$ and has the highest value of $\Pr (p_{jk}\in {\cal I}_{\rm target}|D_{jk})$. 
	\item De-escalation: de-escalate to the dose combination that belongs to $\mathcal{A}_{D_2}$ and has the highest value of $\Pr (p_{jk}\in {\cal I}_{\rm target}|D_{jk})$. 
\end{itemize}
\textbf{Algorithm 4}:  Randomized non-diagonal escalation/de-escalation (\textbf{key4})
\begin{itemize}
	\item Escalation: randomly escalate to the dose combination that belongs to any one of two combination components in $\mathcal{A}_{E_1}$ with randomization probabilities proportional to their respective posterior probabilities $\Pr (p_{jk}\in {\cal I}_{\rm target}|D_{jk})$.
    \item De-escalation: de-escalate to the dose combination that randomly belongs to any one of two combination components in $\mathcal{A}_{D_1}$ with probabilities proportional to their respective posterior probabilities $\Pr (p_{jk}\in {\cal I}_{\rm target}|D_{jk})$.
\end{itemize}
\textbf{Algorithm 5}:  Randomized diagonal  escalation/de-escalation (\textbf{key5})
\begin{itemize}
	\item Escalation: randomly escalate to the dose combination that  belongs to any one of two combination components in $\mathcal{A}_{E_2}$ with randomization probabilities proportional to their respective posterior probabilities $\Pr (p_{jk}\in {\cal I}_{\rm target}|D_{jk})$.
	\item De-escalation: randomly de-escalate to the dose combination that  belongs to any one of two combination components in $\mathcal{A}_{D_2}$ with the randomization probabilities proportional to their respective posterior probabilities $\Pr (p_{jk}\in {\cal I}_{\rm target}|D_{jk})$.
\end{itemize}

For algorithms 1--3 with fixed allocation rules, if there are multiple optimal dose combinations with the same value of $\Pr (p_{jk}\in {\cal I}_{\rm target}|D_{jk})$, we randomly choose one with equal probability.

The trial is completed when the maximum sample size is reached. Given all observed data, we use matrix isotonic regression \cite{dykstra} to obtain the estimate of $p_{jk}$ and select the MTD as the combination with a toxicity estimate that is closest to the target. When there are ties, we randomly choose one as the MTD. For patient safety, during trial conduct, we apply the following overdose control rule: whenever a dose satisfies $\Pr(p_{jk}>\phi|D_{jk})\ge c$ (i.e., the dose is overly toxic),  where $c$ is a prespecified probability cutoff, we eliminate that dose from the trial.  In addition, we impose an early stopping rule: if the lowest dose combination $(1,1)$ is overly toxic, i.e., $\Pr(p_{11}>\phi|D_{11})\ge c$, we terminate the trial early. Throughout the paper, we set $c=0.95$.

\section{Numerical study}
\label{simulation}
\subsection{A random matrix scenario generator}
\label{generator}
The evaluation and comparison of phase I designs are often based on a limited number of prespecified dose--toxicity scenarios, which is prone to favor a specific design intentionally or unintentionally. To avoid that issue, Clertant and O'Quigley \cite{clertant2017semiparametric} proposed a pseudo-uniform algorithm to generate random  toxicity scenarios for more objective and unbiased comparisons for single-agent phase I designs. To the best of our knowledge, there is no accessible algorithm for generating random scenarios for a two-dimensional dose matrix, where the partial order constraint for a two-dimensional toxicity space poses major difficulty.
To fill this gap, we have developed a fast and efficient algorithm to generate random combination scenarios as described below. 

\textbf{Algorithm: a random matrix scenario generator}
\begin{itemize}
	\item[Step 1:] Uniformly choose a dose level $(j,k), 1\le j \le J, 1\le k \le K$ from the $J\times K$ drug-combination space, and set $p_{jk}=\phi$. 
	\item[Step 2:] Given the dose combination $(j,k)$, specify a ``pivotal" path from $p_{11}$ to $p_{1K}$: $p_{11}\rightarrow ... \rightarrow p_{j1}\rightarrow ... \rightarrow p_{jk}\rightarrow ... \rightarrow  p_{jK}\rightarrow ... \rightarrow p_{JK}$.  
The doses in the pivotal path are monotonically ordered in toxicity and partition the $J\times K$ dose matrix into upper block (\textit{UB} above the path) and lower block (\textit{LB} under the path)  matrices.

	\item[Step 3:] Generate the toxicity probabilities for the doses on the pivotal path as follows: generate an ordered uniformly-distributed random sample of toxicity probabilities with length $j+k-2$  from ${\rm Unif}(0,\phi)$ and assign them to $(p_{11},\cdots, p_{j1},\cdots,  p_{jk-1})$, and generate an ordered uniformly-distributed random sample with length 
$J+K-j-k$	from ${\rm Unif}(\phi,p_{\rm max})$  and assign them to ($p_{jk+1},\cdots, p_{jK}, \cdots,  p_{JK}$), where $p_{\rm max}$ is the upper bound for the maximum toxicity rate and is generated from a Beta distribution with mean  $\mu$ and variance $\mu(1-\mu)$. In this paper, we take $\mu = 1-\exp\{-(J\times K)/8\}$, which can satisfactorily provide the upper bound for $p_{JK}$.  

	\item[Step 4:] Generate the toxicity probabilities for the doses located in the \textit{UB} and \textit{LB} as follows:
	\begin{itemize}
		\item For the $UB$, from row $j-1$ to row $1$, sequentially generate $p_{j^\prime k^\prime}\sim {\rm Unif}(p_{j^\prime k^\prime-1},p_{j^\prime+1,k^\prime}), j^\prime=j-1,\ldots,1,k^\prime=2,\ldots,K.$
		
		\item For the $LB$, from row $j+1$ to   row $J$, sequentially generate $p_{j^\prime k^\prime}\sim {\rm Unif}(p_{j^\prime-1,k^\prime-1},p_{j^\prime k^\prime+1}),j^\prime=j+1,\ldots,J,k^\prime=K-1,\ldots,1.$

	\end{itemize}
\end{itemize}
The resulting toxicity probability matrix meets the requirement of  a partial order for drug combinations. For illustration of the proposed random matrix scenario generator, a concrete example of generating a $3\times 3$ toxicity probability matrix is given in  Appendix \ref{randomeg}.  Under the above algorithm, one dose has the toxicity probability equal to the target $\phi$ (i.e., step 1); however, as the toxicity probabilities of other doses are randomly generated, some may be fairly close (although not exactly equal) to $\phi$. As a dose with the toxicity probability within $[\phi-\epsilon_1, \phi+\epsilon_2]$ is deemed acceptable (as the MTD), and the proposed algorithm is able to produce multiple MTDs in the randomly generated toxicity probability matrix.

\subsection{Simulation configuration}
\label{setting}
\begin{figure}
	\centering
	\begin{tabular}{cc}
		\includegraphics[scale=0.5]{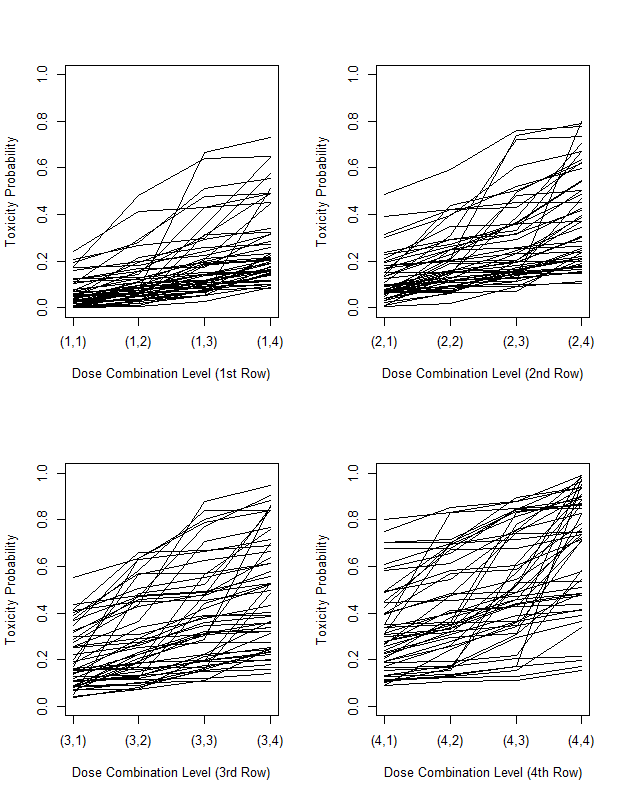}  &
		\includegraphics[scale=0.5]{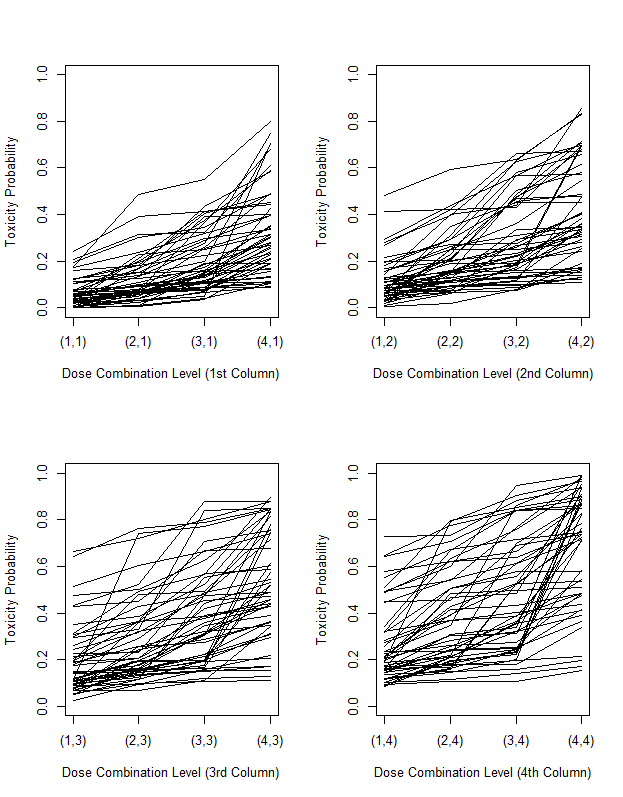}
	\end{tabular}
	\caption{Four rows/columns of 50 randomly selected dose--toxicity curves for the $4\times 4$ drug combination with the target toxicity rate $\phi=0.2$.}
	\label{fig:four-rows-cols-of-4x4}
\end{figure}

\begin{figure}
	\centering
	\includegraphics[scale=0.8]{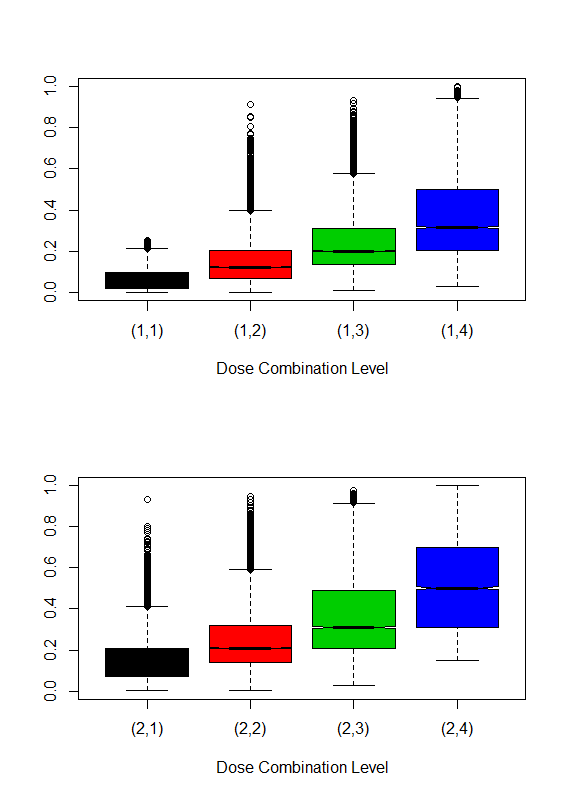}
	\caption{Empirical distribution of the 1000 simulated toxicity probabilities by dose level of the  $2\times 4$ drug combination with the target toxicity rate $\phi=0.2$.}
	\label{fig:sc-dist}
 
\end{figure}

We conduct simulation studies to investigate the performance of the drug-combination keyboard designs based on the random matrix scenarios proposed in Section \ref{generator}. We consider two target toxicity rates ($\phi=0.2$ or 0.3), three sizes of dose matrices ($2\times 4$, $3\times 5$ or $4\times 4$), and three values for the total number of MTDs (1, 2 or 3 MTDs). This results in 16 possible configurations, noting that for the $2\times 4$ combinations, we consider 1 or 2 MTDs.  For each simulation configuration, we generate 1000 random scenarios.  For visualization,  given the target toxicity rate $\phi=0.2$, Figure \ref{fig:four-rows-cols-of-4x4} shows 50 randomly selected dose--toxicity curves for each row and each column for the $4\times 4$ drug combination, while Figure \ref{fig:sc-dist} displays the distributions of the toxicity probabilities by dose combination level from the 1000 random scenarios for the  $2\times 4$ drug combination. These values exhibit a variety of dose--toxicity surface shapes and spacings.

We compare the proposed keyboard combination designs to the partial order CRM method (POCRM) \cite{wages2011continual}. For the POCRM, we choose six partial orderings as recommended by Hirakawa et al.\cite{hirakawa2015comparative}, and assign an equal prior probability to them. The skeleton values are generated according to the algorithm of Lee and Cheung \cite{cheung-book} using the \textit{getprior} function in the R package ``\textit{dfcrm}''. Specifically, for $2\times 4$ combinations, we use \textit{getprior(0.05,$\phi$,4,8)}; for $3\times 5$ combinations, we use \textit{getprior(0.05,$\phi$,7,15)}; and for $4\times 4$ combinations, we use \textit{getprior(0.05,$\phi$,7,16)}. All simulation results regarding POCRM are obtained using the R package ``\textit{pocrm}''. For the keyboard design, we take $\epsilon_1=\epsilon_2=0.03$ for $\phi=0.2$ and  $\epsilon_1=\epsilon_2=0.05$ for $\phi=0.3$, and the resulting dose escalation/de-escalation rules are shown in Table \ref{escrules}. The maximum sample size   is set at 48 for scenarios with $2\times 4$ dimension and at 60 for scenarios with $3\times 5$ and $4\times 4$ dimensions, where the cohort size in all cases is fixed at one for both designs.  For fair comparisons, we apply the early stopping rule as discussed in Section \ref{combination} to both designs.

\subsection{Simulation results}
\label{metric}
Tables \ref{tab:2x4} - \ref{tab:4x4} present simulation results for the $2\times 4$, $3\times 5$ and $4\times 4$ drug combinations, respectively. For a comprehensive comparison, we calculate the following performance metrics across each 1000 random scenarios for POCRM and the keyboard design. 
 
{\bf \textit{MTD selection}}:
 The percentage of correct selection (PCS) is defined as the percentage of simulated trials in which the dose combination selected as the MTD has a true toxicity probability that lies
within the interval $[\phi-\epsilon_1,\phi+\epsilon_2]$. Such a metric quantifies the identification accuracy of a design. In all cases, all keyboard combination designs outperform the POCRM in terms of PCS. For example, when  $\phi=0.2$ with two MTDs in the $3\times5$ combination trials (Table  \ref{tab:3x5}), the PCS of POCRM is 25.1\%, while key1 to key5 have 8.0\%, 7.8\%, 7.4\%, 7.8\% and 7.0\% more chances to  identify the correct MTDs, respectively.  The PCS difference among five versions of the keyboard combination design is not significant.

{\bf \textit{Patient allocation}}: The percentage of correct assignment (PCA) is defined as the  percentage of patients  who are assigned to a dose combination with a true toxicity probability that lies within the interval $[\phi-\epsilon_1,\phi+\epsilon_2]$. Such a metric quantifies the treatment efficiency of a design.  In general, all five versions of the keyboard combination designs have larger PCAs than the POCRM. However, the keyboard designs with fixed allocation rules perform slightly better than those with random allocation rules on average. For example, when $\phi=0.3$ with three MTDs in the $3\times5$ combination trials (Table  \ref{tab:3x5}), key1 to key3 have respective values of 32.12\%, 30.72\%, and 29.68\%, while key4 and key5 have respective values of 27.36\% and 25.97\%.

{\bf\textit{Overdose and underdose control}}:  The percentage of overdose assignment is defined as the percentage of patients who are assigned to a dose combination with a true toxicity probability that is larger than $\phi+\epsilon_2$. The percentage of underdose assignment is defined as the percentage of patients who are assigned to a dose combination with a true toxicity probability that is less than $\phi-\epsilon_1$. Such two metrics respectively quantify the safety and conservativeness of a design. For overdosing, in general, the POCRM allocates fewer patients to doses above the MTD than the keyboard designs. However, for underdosing, the POCRM tends to treat more patients at doses below the MTD.  
 
{\bf \textit{Incoherent index}}: Long-memory incoherence in the escalation percentage is defined as the percentage of dose escalation with  $\hat{p}_{jk}>\phi$. Such a metric can partly quantify the aggressiveness of a design. According to Table \ref{tab:longterm-noncoherent-dlt}, we can conclude that the POCRM is not long-memory coherent. A specific example showing the incoherence (including both short and long memory) of the POCRM can be found in Appendix  \ref{pocrm}. For example, after patient 7 was treated at dose (1, 4), 1/1 patient had DLT at that dose, and the POCRM escalated the dose to (3, 4). When patient 9 was treated at dose (1, 4), 1/2 patients had DLT, and the POCRM again escalated the dose to (3, 3).  
 
In summary,  the  simulation results show that the keyboard combination designs generally outperform the POCRM in terms of MTD identification accuracy and treatment efficiency. 
Among the five versions of keyboard designs,  the designs (key1--key3)
based on fixed allocation rules usually lead to more patients being 
treated at the MTDs than those (key4--key5) that use  randomization. 
We recommend key1 and key2 for general drug-combination trials. While key2 has larger admissible dose escalation and de-escalation sets, key1 tends to be more efficient yet slightly more aggressive than key2.

\section{Web application}
\label{web}
\begin{figure}[http!]
	\centering
	\includegraphics[width=\textwidth, keepaspectratio]{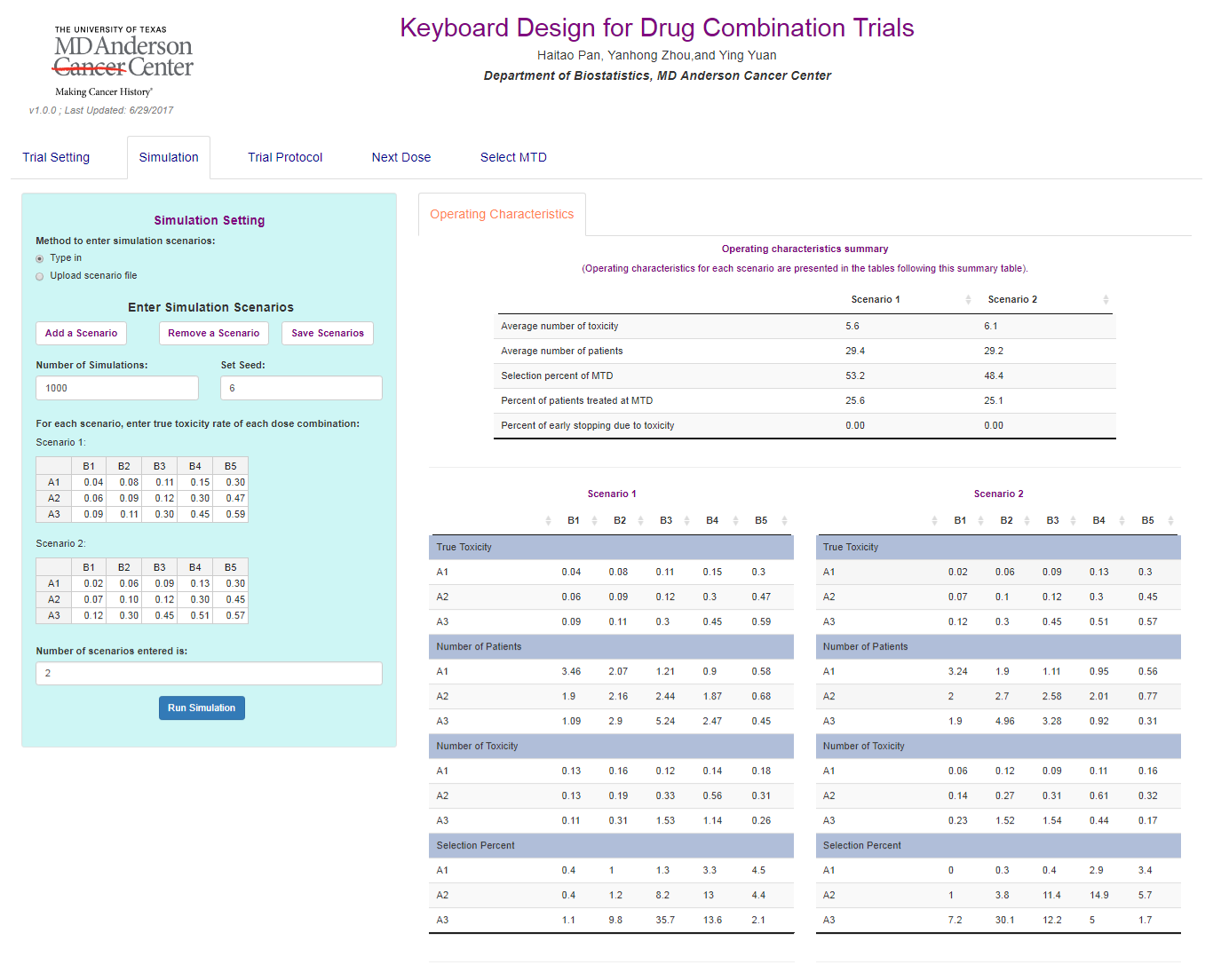}\\
	\caption{Web application for the keyboard combination design.}
	\label{fig:Keyboard-shiny}

\end{figure}

To facilitate the use of the keyboard combination design, we develop an easy-to-implement web application using R Shiny. Figure \ref{fig:Keyboard-shiny} shows the graphical user interface of the application. After users input their design parameters (e.g., the maximum sample size, cohort size, and the dose combination scenarios), the web application generates the operating characteristics of the keyboard combination design that can be included in the trial protocol. When conducting the trial, users can simply choose the ``Next Dose'' button, and the R Shiny application will recommend  the dose combination to be assigned to the next cohort of patients based on the updated accumulated information. Our web application will be freely available at \url{http://www.trialdesign.org}. A reference manual is also provided at that website for illustration and implementation. 

\section{Conclusion}
\label{conclusion}
The statistical properties of the keyboard design have been comprehensively investigated in this paper. 
In particular, we have shown that the keyboard design possesses
optimal allocation rules such that the posterior expected ``0-1'' loss function can be minimized.
The dose transition is long-memory coherent in the sense that 
when the observed toxicity rate at the current dose is larger (smaller) than the target toxicity rate, the keyboard design prevents the next patients from receiving overly toxic (subtherapeutic) doses. In addition, the keyboard design exhibits a convergence property that is similar to that of the standard interval designs, and its asymptotic patient allocation 
converges to the target dose level. 

We have further proposed a class of two-dimensional keyboard designs for dose finding in phase I drug-combination trials. The satisfactory performance and operating characteristics of the keyboard combination designs have been  demonstrated numerically  based on random combination scenarios. From statistical and clinical viewpoints, the proposed combination designs are simple and easy to understand, and we have also developed an R Shiny application for practitioners to use to easily implement the method. Due to the nature of model-assisted designs, the proposed designs are robust against any arbitrary toxicity surfaces of the drug combinations, like the BOIN combination design \cite{lin2017}. Furthermore, the model-assisted keyboard design, without implementing a start-up phase, has comparable performance (better performance in terms of PCS and PCA) compared with the model-based design such as the POCRM. In conclusion, the proposed keyboard combination design offers a middle ground between the existing algorithm-based designs and fully model-based designs.


\newpage

\newpage

  \newpage

\clearpage
  	\begin{table}
  		\caption{Escalation and de-escalation rules for the keyboard design for up to 16 patients treated at the current dose.\label{escrules}}
  		\begin{center}
  			\begin{tabular}{lcccccccccccccccc}
  				\hline\hline
				& \multicolumn{16}{c}{\textbf{Number of patients treated at the current dose}}   \\ 
\cline{2-17}
  				 & 1 & 2 & 3 & 4 & 5 & 6 & 7 & 8 & 9 & 10 & 11 & 12 & 13 & 14 & 15 & 16  \\ 
  				\hline

				\multicolumn{17}{c}{$\phi=0.2$ with the target key = (0.17, 0.23) } \\
  			
 				Escalate if number of DLTs $\le$ & 0 & 0 & 0 & 0 & 0 & 1 & 1 & 1 & 1 & 1 & 1 & 2 & 2 & 2 & 2 & 2  \\
 				De-escalate if number of DLTs $\ge$ & 1 & 1 & 1 & 1 & 2 & 2 & 2 & 2 & 3 & 3 & 3 & 3 & 3 & 4 & 4 & 4 \\
				\\
				 \multicolumn{17}{c}{$\phi=0.3$ with the target key=(0.25, 0.35)} \\
 				Escalate if number of DLTs $\le$ & 0 & 0 & 0 & 0 & 1 & 1 & 1 & 1 & 2 & 2 & 2 & 2 & 3 & 3 & 3 & 3  \\
 				De-escalate if number of DLTs $\ge$ & 1 & 1 & 2 & 2 & 2 & 3 & 3 & 3 & 4 & 4 & 4 & 5 & 5 & 5 & 6 & 6 \\
  				\hline\hline
  			\end{tabular}	
  		\end{center}
  	\end{table}

\begin{table}[]
	\centering
	\caption{Average performance of the six designs across 1000
		random scenarios for the $2\times 4$ drug combination. }
	\scalebox{0.85}{
		\begin{tabular}{lccccccc}
			\hline\hline
			&  &      \multicolumn{6}{c}{\bf Target toxicity rate $\phi=0.2$}        \\ 
			\cline{3-8}
			&  &      \multicolumn{6}{c}{With one MTD}        \\ 
			\cline{4-7}
			&  & POCRM & Key1 & Key2 & Key3 & Key4 & Key5  \\
			\cline{3-8}
			Correct selection\%  & & 30.66  & 33.71  & 33.38      & 33.18  & 32.93  & 32.67           \\
			Correct assignment \%               & & 20.99  & 26.55  & 25.94      & 25.94  & 23.46  & 22.73         \\
			Overdose assignment\% & & 32.15  & 32.93  & 31.23      & 33.51  & 37.79  & 35.36       \\
			Underdose assignment\% &  &   46.86   &  40.52    & 42.83     &  40.55    &  38.75    &     41.91       \\
			
			&  &      \multicolumn{6}{c}{With two MTDs}        \\ 
			\cline{4-7}
			&  & POCRM & Key1 & Key2 & Key3 & Key4 & Key5  \\
			\cline{3-8}
			Correct selection\%  & & 43.65  & 44.14  & 44.68      & 43.46  & 43.28  & 42.83           \\
			Correct assignment \%               &  & 34.31  & 46.51  & 37.00         & 44.67  & 32.37  & 31.22         \\
			Overdose assignment\% && 25.49  & 28.01  & 26.35      & 25.97  & 34.74  & 32.05       \\
			Underdose assignment\% &  &   40.20    &  25.48    & 36.65     &  29.36    &  32.89    &     36.73       \\

			
			\\
			&  &      \multicolumn{6}{c}{\bf Target toxicity rate $\phi=0.3$}        \\ 
			\cline{3-8}
			&  &      \multicolumn{6}{c}{With one MTD}        \\ 
			\cline{4-7}
			&  & POCRM & Key1 & Key2 & Key3 & Key4 & Key5  \\
			\cline{3-8}
			Correct selection\%  & & 21.48  & 38.44  & 38.38      & 37.76  & 37.13  & 37.44           \\
			Correct assignment \%               &  & 14.71  & 27.22  & 26.60       & 26.37  & 23.88  & 23.59         \\
			Overdose assignment\% & & 33.66  & 38.57  & 34.87      & 39.52  & 40.71  & 36.93       \\
			Underdose assignment\% &  & 41.35  & 38.63  & 42.17      & 38.93 & 39.25  & 42.03      \\
			
			&  &      \multicolumn{6}{c}{With two MTDs}        \\ 
			\cline{4-7}
			&  & POCRM & Key1 & Key2 & Key3 & Key4 & Key5  \\
			\cline{3-8}
			Correct selection\%  & & 29.69  & 52.03  & 51.99      & 50.83  & 50.76  & 48.91           \\
			Correct assignment \%               &  & 23.08  & 38.80   & 37.63      & 36.79  & 32.26  & 31.85         \\
			Overdose assignment\% & & 38.36  & 29.80   & 26.35      & 31.34  & 34.04  & 31.29       \\
			Underdose assignment\% &  &   38.56    &  31.40    & 36.02     &  31.87    &  36.86    &     36.90       \\
			
			\hline\hline     
		\end{tabular}
	}
	\label{tab:2x4}
\end{table}

\newpage

\begin{table}[]
	\centering
	\caption{Average performance of the six designs across 1000
		random scenarios for the $3\times 5$ drug combination. }
	\scalebox{0.85}{
	\begin{tabular}{lccccccc}
		\hline\hline
		&  &      \multicolumn{6}{c}{\bf Target toxicity rate $\phi=0.2$}        \\ 
		\cline{3-8}
				&  &      \multicolumn{6}{c}{With one MTD}        \\ 
		\cline{4-7}
		                           &  & POCRM & Key1 & Key2 & Key3 & Key4 & Key5  \\
		 \cline{3-8}
		Correct selection\%  & & 16.38  & 25.07  & 24.85      & 24.92   & 25.41  & 25.02           \\
		Correct assignment \%               &  & 10.19  & 17.72  & 17.24      & 17.43   & 16.73  & 16.07         \\
		Overdose assignment\% & & 31.29  & 36.63  & 34.47      & 37.65   & 41.37  & 38.06       \\
		Underdose assignment\% &  &   58.52    &  45.65    & 48.29     &  44.92    &  41.90    &   45.87       \\
		
						&  &      \multicolumn{6}{c}{With two MTDs}        \\ 
		\cline{4-7}
		&  & POCRM & Key1 & Key2 & Key3 & Key4 & Key5  \\
		\cline{3-8}
				Correct selection\%  & & 25.10   & 33.14  & 32.96      & 32.45   & 32.91  & 32.16           \\
		Correct assignment \%               &  & 18.49  & 24.45  & 23.63      & 23.26   & 21.85  & 20.64         \\
		Overdose assignment\% & & 27.89  & 34.78  & 32.42      & 36.56  & 41.03  & 37.18       \\
		Underdose assignment\% &  &   53.62   &  40.77    & 43.95     &  40.18    &  37.12    &  42.18       \\

						&  &      \multicolumn{6}{c}{With three MTDs}        \\ 
		\cline{4-7}
		&  & POCRM & Key1 & Key2 & Key3 & Key4 & Key5  \\
		\cline{3-8}
	    Correct selection\%  & & 32.43  & 39.73  & 39.60       & 39.08  & 39.06  & 38.37          \\
		Correct assignment \%               &  & 25.37  & 30.79  & 29.78      & 29.24  & 27.21  & 25.81         \\
		Overdose assignment\% & & 24.68  & 24.48  & 24.13      & 34.55   & 39.67  & 35.71       \\
		Underdose assignment\% &  &   49.95    &  44.73    & 46.09     &  36.21    &  33.12    &     38.48       \\    \\                   
		
		&  &      \multicolumn{6}{c}{\bf Target toxicity rate $\phi=0.3$}        \\ 
        \cline{3-8}
				&  &      \multicolumn{6}{c}{With one MTD}        \\ 
\cline{4-7}
&  & POCRM & Key1 & Key2 & Key3 & Key4 & Key5  \\
\cline{3-8}
Correct selection\%  & & 22.39  & 28.71  & 28.50       & 27.97  & 29.65  & 29.13           \\
Correct assignment \%               &  & 13.93  & 18.32  & 17.74      & 17.47  & 17.19  & 16.41         \\
Overdose assignment\% & & 33.66  & 38.57  & 34.87      & 39.52  & 40.71  & 36.93       \\
Underdose assignment\% &  &   52.41    &  43.11    & 47.39     &  43.01    &  42.10    &  46.66       \\

&  &      \multicolumn{6}{c}{With two MTDs}        \\ 
\cline{4-7}
&  & POCRM & Key1 & Key2 & Key3 & Key4 & Key5  \\
\cline{3-8}
Correct selection\%  & & 29.03  & 38.55  & 38.53      & 37.35  & 37.07  & 37.97           \\
Correct assignment \%               &  & 20.43  & 25.21  & 24.24      & 23.43  & 20.89  & 20.72         \\
Overdose assignment\% & & 28.33  & 37.19  & 33.03      & 39.00     & 40.32  & 35.80       \\
Underdose assignment\% &  &   51.24    &  37.60    & 42.73     &  37.57    &  38.79    & 43.48       \\

&  &      \multicolumn{6}{c}{With three MTDs}        \\ 
\cline{4-7}
&  & POCRM & Key1 & Key2 & Key3 & Key4 & Key5  \\
\cline{3-8}
Correct selection\%  & & 36.38  & 46.92  & 46.68      & 45.38  & 46.17  & 45.69           \\
Correct assignment \%               &  & 27.26  & 32.12  & 30.72      & 29.68  & 27.36  & 25.97         \\
Overdose assignment\% & & 25.61  & 34.38  & 30.24      & 36.69  & 38.39  & 33.66       \\
Underdose assignment\% &  &   47.13    &  33.50    & 39.04     &  33.63    &  34.25    &     36.55             \\                   
		\hline\hline     
	\end{tabular}
    }
\label{tab:3x5}
\end{table}

\newpage

\begin{table}[]
	\centering
	\caption{Average performance of the six designs across 1000
		random scenarios for the $4\times 4$ drug combination. }
	\scalebox{0.85}{
		\begin{tabular}{lccccccc}
			\hline\hline
			&  &      \multicolumn{6}{c}{\bf Target toxicity rate $\phi=0.2$}        \\ 
			\cline{3-8}
			&  &      \multicolumn{6}{c}{With one MTD}        \\ 
			\cline{4-7}
			&  & POCRM & Key1 & Key2 & Key3 & Key4 & Key5  \\
			\cline{3-8}
			Correct selection\%  & & 16.21  & 24.71  & 24.45      & 24.69  & 24.89  & 24.31           \\
			Correct assignment \%               &  & 9.12   & 17.07  & 16.54      & 16.81  & 16.10   & 15.37         \\
			Overdose assignment\% & & 20.74  & 38.36  & 36.06      & 39.14  & 42.21  & 38.68       \\
			Underdose assignment\% &  &   70.14    &  44.57    & 47.40     &  44.05    &  41.69    &     45.95       \\
			
			&  &      \multicolumn{6}{c}{With two MTDs}        \\ 
			\cline{4-7}
			&  & POCRM & Key1 & Key2 & Key3 & Key4 & Key5  \\
			\cline{3-8}
			Correct selection\%  & & 25.21  & 32.93  & 32.75      & 32.34  & 32.36  & 31.64           \\
			Correct assignment \%               &  & 15.70   & 23.62  & 22.74      & 22.42  & 21.01  & 19.71         \\
			Overdose assignment\% & & 15.34  & 35.00     & 32.45      & 36.59  & 41.07  & 37.14       \\
			Underdose assignment\% &  &   68.96    &  41.38    & 44.81     &  40.99    &  37.92    &     43.15       \\

			&  &      \multicolumn{6}{c}{With three MTDs}        \\ 
			\cline{4-7}
			&  & POCRM & Key1 & Key2 & Key3 & Key4 & Key5  \\
			\cline{3-8}
			Correct selection\%  & & 34.74  & 39.15  & 39.05     & 38.44  & 38.01  & 37.51          \\
			Correct assignment \%               &  & 22.73  & 29.57  & 28.53      & 27.04  & 25.92  & 24.41         \\
			Overdose assignment\% & & 12.83  & 34.08  & 31.18     & 35.63  & 41.08  & 36.95       \\
			Underdose assignment\% &  &   64.44    &  36.35    & 40.29     &  37.33    &  33.00    &     38.64       \\

			\\
			&  &      \multicolumn{6}{c}{{\bf Target toxicity rate $\phi=0.3$}}        \\ 
			\cline{3-8}
			&  &      \multicolumn{6}{c}{With one MTD}        \\ 
			\cline{4-7}
			&  & POCRM & Key1 & Key2 & Key3 & Key4 & Key5  \\
			\cline{3-8}
			Correct selection\%  & & 21.55  & 28.27  & 27.93      & 27.67  & 28.70   & 27.99           \\
			Correct assignment \%               &  & 12.84  & 17.70   & 17.02      & 16.85  & 16.45  & 15.67         \\
			Overdose assignment\% & & 23.76  & 40.31  & 36.21      & 41.47  & 41.39  & 37.41      \\
			Underdose assignment\% &  &   63.40    &  41.99    & 46.77     &  41.68    &  42.16    &     46.92       \\
			
			&  &      \multicolumn{6}{c}{With two MTDs}        \\ 
			\cline{4-7}
			&  & POCRM & Key1 & Key2 & Key3 & Key4 & Key5  \\
			\cline{3-8}
			Correct selection\%  & & 21.72  & 37.86  & 37.44      & 36.54  & 37.47  & 36.59           \\
			Correct assignment \%               &  & 12.96  & 24.31  & 23.13      & 22.30   & 20.95  & 19.68         \\
			Overdose assignment\% & & 23.69  & 37.78  & 33.26      & 39.63  & 40.36  & 35.63       \\
			Underdose assignment\% &  &   63.35    &  37.91    & 43.61     &  38.07    &  38.69    &     44.69       \\
			
			&  &      \multicolumn{6}{c}{With three MTDs}        \\ 
			\cline{4-7}
			&  & POCRM & Key1 & Key2 & Key3 & Key4 & Key5  \\
			\cline{3-8}
			Correct selection\%  & & 39.93  & 45.35  & 45.12      & 43.87  & 44.60   & 43.86           \\
			Correct assignment \%               &  & 26.06  & 30.33  & 29.02      & 27.79  & 25.82  & 24.42        \\
			Overdose assignment\% & & 14.09  & 36.34  & 31.54      & 38.61  & 39.55  & 34.46       \\
			Underdose assignment\% &  &   59.85    &  33.33    & 39.44     &  33.60    &  34.63    &     41.12             \\                   
			\hline\hline     
		\end{tabular}
	}
	\label{tab:4x4}
\end{table}

\newpage

\newpage
\begin{table}[]
	\caption{Percentage of occurring long-memory incoherence in escalation  for the POCRM among 1000 simulated trials.}
	\begin{center}
\begin{tabular}{ccccc}
	\hline
  Drug combination  & Target rate	& One MTD & Two MTDs  & Three MTDs  \\
  	      \hline
	$2\times 4$     & $0.2$    & 0.53 & 0.61 & NA      \\
                       & $0.3$       & 0.92 &0.99 & NA       \\
	$3\times 5$      & $0.2$    & 1.72 & 1.28 & 1.35     \\
	                   & $0.3$       & 2.32 & 2.62 & 2.76       \\

    $4\times 4$      & $0.2$    & 3.74 & 3.65 & 3.72      \\
                       & $0.3$       & 3.09 & 3.25 & 3.30       \\
	\hline
\end{tabular}
\end{center}
	\label{tab:longterm-noncoherent-dlt} 
	
	{\footnotesize Note: ``NA'' denotes not applicable. }
\end{table}

\newpage
\clearpage

%

\newpage
\appendix
\section*{Appendix}
\label{appendix}
\section{Lemma A: Simplified rule for the keyboard designs}
\label{lemma}
The keyboard designs introduced in Section \ref{method}
adopt $K$ keys $\mathcal{I}_1,\ldots,\mathcal{I}_K$ for decision making. In this Lemma, we show that the decision-making procedure of the keyboard designs can be further simplified to designs using only three keys.

\textbf{Lemma A.} The keyboard designs essentially determine which one of the following keys has the largest posterior probability
\[
\mathcal{K}_{1}:p_{d}\in(\phi-2\epsilon_1-\epsilon_2,\phi-\epsilon_1),\quad \mathcal{K}_{0}:p_{d}\in(\phi-\epsilon_1,\phi+\epsilon_2),\quad \mathcal{K}_{2}:p_{d}\in(\phi+\epsilon_2,\phi+\epsilon_1+2\epsilon_2),
\]
where $\phi+\epsilon_1+2\epsilon_2<1$ and $\phi-2\epsilon_1-\epsilon_2>0$, and the prior $p_{d}\sim{\rm Unif}(0,1)$.
The dose assignment decisions in the keyboard design are then
\begin{itemize}
	\item Escalate the dose if $1=\underset{i\in\{0,1,2\}}{\rm argmax}\{\Pr(\mathcal{K}_{i}\mid D_d)\}$;
	\item De-escalate the dose if $2=\underset{i\in\{0,1,2\}}{\rm argmax}\{\Pr(\mathcal{K}_{i}\mid D_d)\}$;
	\item Retain the current dose if $0=\underset{i\in\{0,1,2\}}{\rm argmax}\{\Pr(\mathcal{K}_{i}\mid D_d)\}$.
\end{itemize}
\begin{proof}
Let $k^\dag$ denote the index of the target key among the $K$ prespecified keys, i.e., $\mathcal{I}_{k^\dag}=\mathcal{I}_{\rm target}=\mathcal{K}_{0}$.
The posterior probability of event $\mathcal{K}_0$ (or $\mathcal{I}_{k^\dag}$) is given by 
\begin{eqnarray*}
	\Pr(\mathcal{K}_{0}\mid D_d) & = & \int1\{p_{d}\in \mathcal{K}_{0}\}f(p_{d}\mid D_d){\rm d}p_{d}\\
	& \propto & \int_{\phi-\epsilon_1}^{\phi+\epsilon_2}p_{d}^{y_{d}}(1-p_{d})^{n_{d}-y_{d}}{\rm d}p_{d}\\
	&  \propto  & \xi_{0}^{y_{d}}(1-\xi_{0})^{n_{d}-y_{d}}\equiv g(\xi_{0})
\end{eqnarray*}
where the last equality is due to the mean value theorem for integrals and $\xi_{0}$ is a toxicity probability   that lies within $(\phi-\epsilon_1,\phi+\epsilon_2)$. Similarly, the posterior probabilities of 
events $\mathcal{K}_1$ and $\mathcal{K}_2$ can be represented as $\Pr(\mathcal{K}_{1}\mid D_d)\propto g(\xi_{1})$
and $\Pr(\mathcal{K}_{2}\mid D_d)\propto g(\xi_{2})$, respectively, where 
$\xi_1\in (\phi-2\epsilon_1-\epsilon_2,\phi-\epsilon_1)$ and $\xi_2 \in (\phi+\epsilon_2,\phi+\epsilon_1+2\epsilon_2)$,
$\xi_1<\xi_0<\xi_2$. 

Three cases are considered.
\begin{itemize}
\item For dose retainment, 
\begin{itemize}
\item if $0=\underset{i\in\{0,1,2\}}{\rm argmax}\{\Pr(\mathcal{K}_{i}\mid D_d)\}$, then $g(\xi_0)\geq \max(g(\xi_1),g(\xi_2))$. Given 
a key $k$ with ${\cal I}_{k}\prec {\cal I}_{k^\dag-1} (\mbox{or }\mathcal{K}_1)$, we have $\Pr(p_d\in\mathcal{I}_{k}\mid D_{d})\propto g(\xi^{\prime})$, where $\xi^{\prime}\in\mathcal{I}_{k}$ and $\xi^{\prime}<\xi_{1}$. Then, according to the unimodality of the binomial likelihood, it is easy to obtain that $g(\xi^{\prime})<g(\xi_1)<g(\xi_{0})$ and thus $\Pr(\mathcal{I}_{k}\mid D_d)<\Pr(\mathcal{I}_{k^\dag}\mid D_d)$.
Similarly, given a key $k$ with ${\cal I}_{k}\succ {\cal I}_{k^\dag+1} (\mbox{or }\mathcal{K}_2)$, we can also show that $\Pr(\mathcal{I}_{k}\mid D_d)<\Pr(\mathcal{I}_{k^\dag}\mid D_d)$. Therefore, the keyboard designs based on three keys imply those based on $K$ keys in terms of dose retainment. 

\item if ${\cal I}_{\text{max}}\equiv {\cal I}_{\text{\rm target}}$, then we immediately obtain that $0=\underset{i\in\{0,1,2\}}{\rm argmax}\{\Pr(\mathcal{K}_{i}\mid D_d)\}$. Therefore, the keyboard designs based on $K$ keys also imply those based on three keys in terms of dose retainment. 
\end{itemize}

\item For dose escalation, 
\begin{itemize}
\item  if $1=\underset{i\in\{0,1,2\}}{\rm argmax}\{\Pr(\mathcal{K}_{i}\mid D_d)\}$, then $g(\xi_1) > g(\xi_0)> g(\xi_2)$ which is due to the unimodality of the binomial likelihood, leading to  
$\mathcal{I}_{\rm max}\prec {\cal I}_{\text{\rm target}}$. Therefore, the keyboard designs based on three keys imply those based on $K$ keys in terms of dose escalation.

\item if ${\cal I}_{\text{max}}\prec {\cal I}_{\text{\rm target}}$, then using the unimodality of the binomial likelihood again, we obtain that 
$g(\xi_2) < g(\xi_0)< g(\xi_1)< g(\xi^\prime)$, where 
$\Pr(\mathcal{I}_{\rm max}\mid D_{d})\propto g(\xi^{\prime})$,
leading to 
$1=\underset{i\in\{0,1,2\}}{\rm argmax}\{\Pr(\mathcal{K}_{i}\mid D_d)\}$. Therefore, the keyboard designs based on $K$ keys also imply those based on three keys in terms of dose escalation. 
\end{itemize}

\item For dose de-escalation, it can be shown similarly that  the keyboard designs based on three keys are equivalent to  those based on $K$ keys in terms of dose de-escalation. 

\end{itemize}

\end{proof}

\section{Proof of Theorem 2 - Coherence}
\label{coherence}
\begin{proof}
According to Lemma \ref{lemma}, the decision of the keyboard designs is solely based on the maximum value among $\{g(\xi_1),g(\xi_0),g(\xi_2)\}$, or equivalently, among $\{\Pr(\mathcal{K}_{1}\mid D_d)\},\{\Pr(\mathcal{K}_{0}\mid D_d)\},\{\Pr(\mathcal{K}_{2}\mid D_d)\}$.
Since  the binomial likelihood $g(p_d)=p_{d}^{y_{d}}(1-p_{d})^{n_{d}-y_{d}}$
is maximized at $\hat{p}_{d}=y_{d}/n_{d}$, we have the following possible scenarios.
\begin{itemize}
	\item If $\hat{p}_{d}\leq\phi-\epsilon_1$, according to the unimodality of the
	binomial likelihood, we always have $g(\xi_{2})< g(\xi_{0})$,
	which indicates that $\Pr(\mathcal{K}_{2}\mid D_d)<\Pr(\mathcal{K}_{0}\mid D_d)$,
	then the chance for dose de-escalation is zero. 
	\item If $\hat{p}_{d}\geq\phi+\epsilon_2$, according to the unimodality of the
	binomial likelihood, we always have $g(\xi_{1})< g(\xi_{0})$,
	which indicates that $\Pr(\mathcal{K}_{1}\mid D_d)<\Pr(\mathcal{K}_{0}\mid D_d)$,
	then the chance for dose escalation is zero. 
	\item If $\phi\leq \hat{p}_d<\phi+\epsilon_2$, we define the following difference function, 
	\begin{eqnarray*}
d(\epsilon_{1},\epsilon_{2}) & = & \Pr(\mathcal{K}_{0}\mid D_{d})-\Pr(\mathcal{K}_{1}\mid D_{d})\\
 & \propto & \int_{\phi-\epsilon_{1}}^{\phi+\epsilon_{2}}g(p_{d}){\rm d}p_{d}-\int_{\phi-2\epsilon_{1}-\epsilon_{2}}^{\phi-\epsilon_{1}}g(p_{d}){\rm d}p_{d}.
\end{eqnarray*}
Differentiating $d(\epsilon_{1},\epsilon_{2})$ with respect to $\epsilon_1$, we have \[
\frac{\partial d(\epsilon_{1},\epsilon_{2})}{\partial\epsilon_{1}}=2g(\phi-\epsilon_{1})-2g(\phi-2\epsilon_{1}-\epsilon_{2})>0,
\]
as the mode of $g(p_d)$ is $\hat{p}_d$ and $\hat{p}_{d}\geq\phi-\epsilon_1$. Therefore, $d(\epsilon_{1},\epsilon_{2})$ is an increasing function of $\epsilon_1$, and we only need to consider the case when $\epsilon_1=0$. That is, if we can show that $d(0,\epsilon_{2})>0$ then based on the monotonicity of $d(\epsilon_{1},\epsilon_{2})$, we have $d(\epsilon_{1},\epsilon_{2})>d(0,\epsilon_{2})>0$. 

When $\epsilon_1=0$, the difference function reduces to 
\[
d(0,\epsilon_{2})\propto\int_{\phi}^{\phi+\epsilon_{2}}g(p_{d}){\rm d}p_{d}-\int_{\phi-\epsilon_{2}}^{\phi}g(p_{d}){\rm d}p_{d}.
\]
Again, differentiating $d(0,\epsilon_{2})$ with respect to $\epsilon_2$, we have
\[
\frac{\partial d(0,\epsilon_{2})}{\partial\epsilon_{2}}\propto g(\phi+\epsilon_{2})-g(\phi-\epsilon_{2}).
\]
Denote the log-likelihood ratio of $\phi+\epsilon_2$ versus $\phi-\epsilon_2$ as 
\begin{eqnarray*}
l(y_{d},n_{d},\epsilon_2) & = & \ln\frac{g(\phi+\epsilon_{2})}{g(\phi-\epsilon_{2})}\\
 & = & y_{d}\ln(\phi+\epsilon_{2})+(n_{d}-y_{d})\ln(1-\phi-\epsilon_{2})\\
 &  & -y_{d}\ln(\phi-\epsilon_{2})-(n_{d}-y_{d})\ln(1-\phi+\epsilon_{2}),
\end{eqnarray*}
which is an increasing function with respect to $y_d$.
Since $\hat{p}_{d}\in(\phi,\phi+\epsilon_2)$ and $y_d$ is a integer value, it suffices to show that 
$d(0,\epsilon_{2})>0$ with $D_d=(y_d+1,n_d)$, $\hat{p}_{d}=(y_{d}+1)/n_{d}$, and $\phi=y_d/n_d$. 

Next, we consider two cases:
\begin{itemize}
\item[(a)] When $\phi<0.5$, taking the derivative of $l(y_{d},n_{d},\epsilon_2)$ with respect to $\epsilon_2$, we have
\begin{eqnarray*}
\frac{\partial l(y_{d}+1,n_{d},\epsilon_{2})}{\partial\epsilon_{2}} & = & \frac{y_{d}+1}{\phi+\epsilon_{2}}-\frac{n_{d}-y_{d}-1}{1-\phi-\epsilon_{2}}+\frac{y_{d}+1}{\phi-\epsilon_{2}}-\frac{n_{d}-y_{d}-1}{1-\phi+\epsilon_{2}}\\
 & = & \frac{2\epsilon_{2}^{2}(n_{d}-2\phi n_{d}-1)+2\phi(1-\phi)}{(\phi+\epsilon_{2})(1-\phi-\epsilon_{2})(\phi-\epsilon_{2})(1-\phi+\epsilon_{2})}.
 \end{eqnarray*}
Since $\phi-\epsilon_2>0$, we have 
$$\frac{\partial l(y_{d}+1,n_{d},\epsilon_{2})}{\partial\epsilon_{2}}>\frac{\partial l(y_{d}+1,1,\phi)}{\partial\epsilon_{2}}=\frac{2(\phi-\phi^{2}-2\phi^{3})}{(\phi+\epsilon_{2})(1-\phi-\epsilon_{2})(\phi-\epsilon_{2})(1-\phi+\epsilon_{2})}\geq0.$$
Therefore, $l(y_{d}+1,n_{d},\epsilon_2)$ is an increasing function of $\epsilon_2$, which implies that $l(y_{d}+1,n_{d},\epsilon_2)>l(y_{d}+1,n_{d},0)=1$.
As a result,
$$d(\epsilon_{1},\epsilon_{2})>d(0,\epsilon_{2})>d(0,0)=0.$$

\item[(b)] When $\phi \geq0.5$, we have $\phi+\epsilon_2 \leq 1$. Since $d(0,\epsilon_2)$ is a unimodal function with respect to $\epsilon_2$, we just need to focus on the two end points $\epsilon_2=0$ and $\epsilon_2=1-\phi$.  If $\epsilon_2=0$, 
it is easy to observe that $d(0,0)=0$. If $\epsilon_2=1-\phi$, 
then 
$$d(0,1-\phi)	\propto	\int_{\phi}^{1}g(p_{d}){\rm d}p_{d}-\int_{2\phi-1}^{\phi}g(p_{d}){\rm d}p_{d},$$
and 
$$\frac{\partial d(0,1-\phi)}{\partial\phi}\propto2g(2\phi-1)-2g(\phi)\leq0,$$
given that $\phi \geq0.5$,  $y_d=n_d\phi$, 
and $g(\cdot)$ is evaluated at $D_d=(y_d+1,n_d)$. As a result, 
$d(0,1-\phi)$ is a decreasing function of $\phi$, and 
$$d(\epsilon_{1},\epsilon_{2})>d(0,\epsilon_{2})> \min(d(0,0), d(0,1-\phi)) \geq d(0,0)=0.$$
\end{itemize}
In summary, $d(\epsilon_{1},\epsilon_{2})  =  \Pr(\mathcal{K}_{0}\mid D_{d})-\Pr(\mathcal{K}_{1}\mid D_{d})>0$, the chance for dose escalation is zero.

	\item If $\phi-\epsilon_1< \hat{p}_d<\phi$, based on a mimicking approach of the case in which $\phi\leq \hat{p}_d<\phi+\epsilon_2$, it can be shown that  $\Pr(\mathcal{K}_{2}\mid D_d)<\Pr(\mathcal{K}_{0}\mid D_d)$ and the chance for dose de-escalation is zero.
	
\end{itemize}
Based on the above argument, the keyboard design is long-memory coherent. 
\end{proof}
\bigskip

\section{Proof of Theorem 3 - Convergence}
\label{convergence}
\begin{proof}
	In what follows, we provide a heuristic proof of the convergence of the keyboard designs.  As the sample size increases, it is well known that the posterior distribution of $p_d$ under the beta-binomial model asymptotically converges to a normal distribution:
	\begin{align}
		p_d \sim N(\hat{p}_d,[I(D_d)]^{-1})
	\end{align}
	where $\hat{p}_d= {\rm arg~max}_{p_d} f(D_d|p_d)f(p_d)$ and $[I(D_d)]^{-1}=-\{ \frac{\partial^2}{\partial p_d^2 } \log[f(D_d|p_d)f(p_d)] \}|_{p_d=\hat{p_d}}$. Here, $f(D_d|p_d)$ is the likelihood function of the data at the dose level $d$ and $f(p_d)$ is the corresponding prior density, $p_d\sim {\rm Unif}(0,1)$.
	
	From the above, we see that the mode of the asymptotic distribution is a consistent estimate of $p_d$. That is, under large samples, the strongest key ${\cal I}_{max}$ converges to the 
key $\mathcal{I}_k$ with $p_d\in \mathcal{I}_k$. Therefore, based on the keyboard design rules,
	\begin{itemize}
		\item if ${\cal I}_{max} \prec {\cal I}_{\text{\rm target}}$, we escalate the dose, and ${\cal I}_{max}^\prime$, the strongest key  of the next dose level, is on the right side of ${\cal I}_{max}$;
		\item if ${\cal I}_{max} \succ {\cal I}_{\text{\rm target}}$, we de-escalate the dose, and ${\cal I}_{max}^\prime$, the strongest key of the next dose level, is on the left side of ${\cal I}_{max}$.
	\end{itemize}
 After a long run, as long as the next dose level hits the dose with $p_d\in {\cal I}_{\text{\rm target}}$, according to the asymptotic posterior distribution of $p_d$, we have $\Pr(\mathcal{I}_{{\rm max}}=\mathcal{I}_{{\rm target}}\mid D_{d})=1$ as $n_d\rightarrow \infty$. 
 As a result, the asymptotic dose movement of the keyboard designs can only settle  on a dose with the toxicity probability $p_d$ within the target key ${\cal I}_{\text{\rm target}}$.

\end{proof}

\section{An example of generating a random $3\times 3$ toxicity probability matrix.}
\label{randomeg}
For illustration of the proposed random scenario generator for drug-combination trials, we consider a $3\times 3$ toxicity probability matrix  with the target toxicity rate $\phi=0.2$,
\[
\begin{bmatrix}
p_{11}       & p_{12} & p_{13}   \\
p_{21}       & p_{22} & p_{23}  \\
p_{31}       & p_{32} & p_{33}  
\end{bmatrix}.
\]
According to the proposed algorithm as described in Section \ref{generator}, 
\begin{itemize}
	\item[Step 1:] Randomly choose an element from the $3\times 3$ matrix with an equal probability (of 1/9) and set it as $\phi$. For instance, dose combination $(2,2)$ is selected, and now, the probability matrix is
	\[
	\begin{bmatrix}
	p_{11}       & p_{12} & p_{13}   \\
	p_{21}       & {\color{blue} p_{22}=0.20} & p_{23}  \\
	p_{31}       & p_{32} & p_{33}  
	\end{bmatrix}
	\]
	\item[Step 2:] Specify the pivotal path, given by $p_{11} \to p_{21} \to p_{22} \to p_{23} \to p_{33}$, as shown below.
		\[
	\begin{bmatrix}
	{\color{blue} p_{11}}       & p_{12} & p_{13}   \\
	{\color{blue} p_{21} }      & {\color{blue}  p_{22}=0.20} & {\color{blue} p_{23}}  \\
	p_{31}       & p_{32} & {\color{blue} p_{33}  }
	\end{bmatrix}
	\]
	\item[Step 3:] Generate the toxicity probabilities for the doses on the pivotal path as follows. Generating ($p_{11},p_{21}$) from ${\rm Unif}(0,0.2)$,  we obtain $p_{11}=0.01$ and $p_{21}=0.15$. As the upper bound of $p_{\rm max}=1-\exp(-3\time3/8)=0.68$, we generate $(p_{23}, p_{33})$ from ${\rm Unif}(0.20,0.68)$ and obtain $p_{23}=0.38$ and $p_{33}=0.55$.  Now, the toxicity probability matrix is
	\[
	\begin{bmatrix}
	{\color{blue} p_{11}=0.01 }      & p_{12} & p_{13}   \\
	{\color{blue} p_{21}=0.15  }     & {\color{blue} p_{22}=0.20 } & {\color{blue}  p_{23}=0.38 }  \\
	p_{31}       &  p_{32} & {\color{blue}  p_{33}=0.55  }
	\end{bmatrix}
	\]
	\item[Step 4:] For the upper block (UB),  $p_{12}$ can be generated from  Unif(1,0.01,0.20)=0.10, and  $p_{13}$ from Unif(1,0.10,0.38)=0.23. Thus, we completed the UB part, and the matrix is
		\[
	\begin{bmatrix}
	{\color{blue} p_{11}=0.01}       & {\color{blue} p_{12}=0.10} & {\color{blue} p_{13}=0.23}   \\
	{\color{blue} p_{21}=0.15}       & {\color{blue} p_{22}=0.20 } & {\color{blue} p_{23}=0.38}  \\
	p_{31}       & p_{32} & {\color{blue} p_{33}=0.55  }
	\end{bmatrix}
	\]
	Similarly, for the lower block (LB),   $p_{32}$ can be generated by Unif(1,0.2,0.55)=0.42 and   $p_{31}$ by Unif(1,0.15,0.42)=0.27. The final matrix is
			\[
			{\color{blue} 
	\begin{bmatrix}
	p_{11}=0.01       & p_{12}=0.10 & {\bf p_{13}=0.23}   \\
	p_{21}=0.15      &  {\bf p_{22}=0.20} & p_{23}=0.38  \\
	p_{31} =0.27      & p_{32}=0.42 & p_{33}=0.55  
	\end{bmatrix}}
	\]
\end{itemize}
If we set $\epsilon_1=\epsilon_2=0.03$, we can see that we generated a random matrix having two MTDs, i.e.,  $p_{13}$ and $p_{22}$.

\section{An example of incoherence for POCRM}
\label{pocrm}
This example illustrates the violation of 
incoherence for POCRM.
We consider a $4\times 4$ matrix setting. The target toxicity rate is 0.30 and the true toxicity probability matrix is 
			\[
			\{(i,j)\}_{i,j=1}^4=
\begin{bmatrix}
0.50& 0.55& 0.60& 0.70  \\
0.15& {\bf 0.30}& 0.50& 0.60  \\
0.10& 0.12& {\bf 0.30}& 0.45  \\
0.06& 0.08& 0.10& 0.15
\end{bmatrix}
\]
Following \cite{wages2011continual}, three orders are prespecified for the POCRM, 
\begin{eqnarray*}
\text{Order 1} & : & (1,1)\rightarrow(1,2)\rightarrow(2,1)\rightarrow(1,3)\rightarrow(2,2)\rightarrow(3,1)\rightarrow(1,4)\rightarrow(2,3)\rightarrow\\
 &  & (3,2)\rightarrow(4,1)\rightarrow(2,4)\rightarrow(3,3)\rightarrow(4,2)\rightarrow(3,4)\rightarrow(4,3)\rightarrow(4,4)\\
\text{Order 2} & : & (1,1)\rightarrow(2,1)\rightarrow(1,2)\rightarrow(1,3)\rightarrow(2,2)\rightarrow(3,1)\rightarrow(4,1)\rightarrow(3,2)\rightarrow\\
 &  & (2,3)\rightarrow(1,4)\rightarrow(2,4)\rightarrow(3,3)\rightarrow(4,2)\rightarrow(4,3)\rightarrow(3,4)\rightarrow(4,4)\\
\text{Order 3} & : & (1,1)\rightarrow(2,1)\rightarrow(1,2)\rightarrow(3,1)\rightarrow(2,2)\rightarrow(1,3)\rightarrow(4,1)\rightarrow(3,2)\rightarrow\\
 &  & (2,3)\rightarrow(1,4)\rightarrow(4,2)\rightarrow(3,3)\rightarrow(2,4)\rightarrow(4,3)\rightarrow(3,4)\rightarrow(4,4)
\end{eqnarray*}
The maximum sample size is 40, and we utilize ``{\it pocrm.sim}'' in the R package ``{\it pocrm}'' 
to simulate one trial based on the random seed of 2.
The other implementation details of the POCRM are exactly the same as those used in the reference manual of ``{\it pocrm}''.

We provide step-by-step movements of POCRM in  Table \ref{tab:dose move}. From the table, we can see that for this simulated trial, there are 4 incoherent dose movements.
 \setcounter{table}{0}
\renewcommand{\thetable}{A\arabic{table}}
\begin{table}[]
\setlength{\tabcolsep}{0.7mm}
{\small
	\caption{Dose movements of the POCRM for a $4\times 4$ combination trial with $\phi=0.3$.}\label{tab:dose move}
\begin{tabular}{ccccccccccccccc}
\hline 
Patient & Dose & DLT & $\hat{p}_{jk}$ & Order & Action & Incoherent &  & Patient & Dose & DLT & $\hat{p}_{jk}$ & Order & Action & Incoherent\tabularnewline
\hline 
1 & $(1,1)$ & 0 & 0/1 & NA & $\mathcal{E}$ & NA &  & 21 & $(4,2)$ & 1 & 1/2 & 3 & $\mathcal{D}$ & N\tabularnewline
2 & $(1,2)$ & 0 & 0/1 & NA & $\mathcal{E}$ & NA &  & 22 & $(2,3)$ & 1 & 1/2 & 3 & $\mathcal{D}$ & N\tabularnewline
3 & $(2,1)$ & 0 & 0/1 & NA & $\mathcal{E}$ & NA &  & 23 & $(4,1)$ & 1 & 1/1 & 1 & $\mathcal{D}$ & N\tabularnewline
4 & $(1,3)$ & 0 & 0/1 & NA & $\mathcal{E}$ & NA &  & 24 & $(2,2)$ & 0 & 0/2 & 1 & $\mathcal{E}$ & N\tabularnewline
5 & $(2,2)$ & 0 & 0/1 & NA & $\mathcal{E}$ & NA &  & 25 & $(3,1)$ & 1 & 1/2 & 1 & $\mathcal{D}$ & N\tabularnewline
6 & $(3,1)$ & 0 & 0/1 & NA & $\mathcal{E}$ & NA &  & 26 & $(1,3)$ & 1 & 1/2 & 1 & $\mathcal{D}$ & N\tabularnewline
7 & $(1,4)$ & 1 & 1/1 & 3 & $\mathcal{E}$ & {\bf \underline{\color{red} Y}} &  & 27 & $(2,1)$ & 0 & 0/2 & 1 & $\mathcal{E}$ & N\tabularnewline
8 & $(3,4)$ & 1 & 1/1 & 3 & $\mathcal{D}$ & N &  & 28 & $(1,3)$ & 0 & 1/3 & 1 & $\mathcal{R}$ & N\tabularnewline
9 & $(1,4)$ & 0 & 1/2 & 3 & $\mathcal{E}$ & {\bf \underline{\color{red} Y}} &  & 29 & $(1,3)$ & 0 & 1/4 & 1 & $\mathcal{E}$ & N\tabularnewline
10 & $(3,3)$ & 0 & 0/1 & 2 & $\mathcal{E}$ & N &  & 30 & $(2,2)$ & 0 & 0/3 & 1 & $\mathcal{R}$ & N\tabularnewline
11 & $(4,3)$ & 0 & 0/1 & 2 & $\mathcal{E}$ & N &  & 31 & $(2,2)$ & 0 & 0/4 & 1 & $\mathcal{E}$ & N\tabularnewline
12 & $(3,4)$ & 1 & 2/2 & 3 & $\mathcal{D}$ & N &  & 32 & $(3,1)$ & 0 & 1/3 & 1 & $\mathcal{R}$ & N\tabularnewline
13 & $(3,3)$ & 0 & 0/2 & 2 & $\mathcal{E}$ & N &  & 33 & $(3,1)$ & 0 & 1/4 & 1 & $\mathcal{E}$ & N\tabularnewline
14 & $(4,3)$ & 1 & 1/2 & 2 & $\mathcal{D}$ & N &  & 34 & $(1,4)$ & 0 & 1/5 & 1 & $\mathcal{R}$ & N\tabularnewline
15 & $(2,4)$ & 1 & 1/1 & 3 & $\mathcal{D}$ & N &  & 35 & $(1,4)$ & 0 & 1/6 & 1 & $\mathcal{E}$ & N\tabularnewline
16 & $(2,3)$ & 0 & 0/1 & 3 & $\mathcal{E}$ & N &  & 36 & $(2,3)$ & 0 & 1/3 & 1 & $\mathcal{R}$ & N\tabularnewline
17 & $(1,4)$ & 0 & 1/3 & 3 & $\mathcal{E}$ & {\bf \underline{\color{red} Y}} &  & 37 & $(2,3)$ & 1 & 2/4 & 1 & $\mathcal{D}$ & N\tabularnewline
18 & $(4,2)$ & 0 & 0/1 & 3 & $\mathcal{E}$ & N &  & 38 & $(1,4)$ & 1 & 2/7 & 1 & $\mathcal{D}$ & {\bf \underline{\color{red} Y}}\tabularnewline
19 & $(3,3)$ & 1 & 1/3 & 3 & $\mathcal{D}$ & N &  & 39 & $(3,1)$ & 0 & 1/4 & 3 & $\mathcal{E}$ & N\tabularnewline
20 & $(1,4)$ & 0 & 1/4 & 3 & $\mathcal{E}$ & N &  & 40 & $(4,1)$ & 0 & 1/2 & 3 & $\mathcal{R}$ & N\tabularnewline
\hline 
\end{tabular}
}

{\footnotesize The first six patients were treated in the start-up phase. Actions $\mathcal{E},\mathcal{R}, \mathcal{D}$ denote the decisions of dose escalation, retainment, and de-escalation, respectively. The incoherence indicator is calculated based on the estimated toxicity order. }
\end{table}

\end{document}